\documentclass[prd,aps,floats,floatfix,nofootinbib]{revtex4}

\usepackage{amsmath}
\usepackage{amsfonts}
\usepackage{graphicx}
\usepackage{graphicx}
\usepackage{dcolumn}
\usepackage{bm}
\usepackage[dvips]{color}
\usepackage{slashed}
\newcommand{\beq}{\begin{equation}}
\newcommand{\eeq}{\end{equation}}
\begin{document}

\title{The Limits of Custodial Symmetry\footnote{Speaker at conference: R. Sekhar Chivukula.
This report is a shortened version of previously published
work \protect\cite{SekharChivukula:2009if}.}}

\author{R. Sekhar Chivukula$^a$, Roshan Foadi$^b$, and Elizabeth H. Simmons$^c$}

\address{Department of Physics and Astronomy,\\
Michigan State University, East Lansing, MI 48824, USA\\
$^a$sekhar@msu.edu, $^b$foadiros@msu.edu, $^c$esimmons@msu.edu}

\author{Stefano Di Chiara}

\address{CP3-Origins, Campusvej 55, DK-5230 Odense M, Denmark.\\
dichiara@cp3.sdu.dk}

\begin{abstract}
We introduce a toy model implementing the proposal of using a custodial symmetry to protect the $Z b_L \bar{b}_L$ coupling from large corrections.  This ``doublet-extended standard model" adds a weak doublet of fermions (including a heavy partner of the top quark) to the particle content of the standard model in order to implement an $O(4) \times U(1)_X  \sim SU(2)_L \times SU(2)_R \times P_{LR} \times U(1)_X$ symmetry in the top-quark mass generating sector.  This symmetry is softly broken to the gauged $SU(2)_L \times U(1)_Y$ electroweak symmetry by a Dirac mass $M$ for the new doublet; adjusting the value of $M$ allows us to explore the range of possibilities between the $O(4)$-symmetric ($M \to 0$) and standard-model-like ($M \to \infty$) limits. 
\end{abstract}

%\keywords{Style file; \LaTeX; Proceedings; World Scientific Publishing.}

\maketitle

\section{Introduction}

Agashe \cite{Agashe:2006at} et al.  have shown that the constraints on beyond the standard model physics related to the $Zb_L\bar{b}_L$ coupling can, in principle, be loosened if the global $SU(2)_L \times SU(2)_R$ symmetry of the electroweak symmetry breaking sector is actually a subgroup of a larger global symmetry  of both the symmetry breaking and top quark mass generating sectors of the theory. In particular, they propose that these interactions preserve an  $O(4)\sim SU(2)_L\times SU(2)_R\times P_{LR}$ symmetry, where $P_{LR}$ is a parity interchanging $L\leftrightarrow R$. The $O(4)$ symmetry is then spontaneously broken to  $O(3)\sim SU(2)_V\times P_{LR}$,  breaking the elecroweak interactions but
protecting $g_{Lb}$ from radiative corrections, so long as the left-handed bottom quark is a $P_{LR}$ eigenstate.

In this talk we report on the construction of the simplest $O(4)$-symmetric extension of the SM.  For reasons that will shortly become clear, we call this model the doublet-extended standard model or DESM.  Because the DESM is minimal, it displays the essential ingredients protecting $g_{Lb}$ without the burden of additional states, interactions, or symmetry patterns that might otherwise obscure the role played by custodial $O(3)$.  Because it is concrete, it also enables us to explore how the new symmetry impacts the model's ability to conform with the constraints imposed by other precision electroweak data.

In our model, all operators of dimension-4 in the Higgs potential and the sector generating the top quark mass respect a global $O(4)\times U(1)_X$ symmetry; the $U(1)_X$ enables the SM-like fermions to obtain the appropriate electric charges and hypercharges.  In addition to the particle content of the SM, we introduce a new weak doublet of Dirac fermions, $\Psi=(\Theta,T^\prime)$, and combine $\Psi_L$  with the left-handed top-bottom doublet $(t^\prime_L, b_L)$ to form a $(2,2^*)$ under the global $SU(2)_L\times SU(2)_R$ symmetry.  The $b_L$ state is thereby endowed with identical charges under the two global $SU(2)$ groups, $T_L^3=T_R^3$, making it a parity eigenstate, as desired.   We also find that the $T'$ mixes with $t^\prime$ to form a SM-like top quark and a heavy partner.  The $O(4) \times U(1)_X$ symmetric Yukawa interaction can, of course, be extended to the bottom quark and the remaining electroweak doublets, by adding further spectator fermions;  here we focus exclusively on the partners of the top quark since they give the dominant contribution to $g_{Lb}$.

To enable electroweak symmetry breaking and fermion mass generation to proceed, the global symmetry is explicitly broken to $SU(2)_L\times U(1)_Y$ by a dimension-three Dirac mass $M$ for $\Psi$. As $M\to\infty$ the ordinary SM top Yukawa interaction is recovered; as $M\to 0$ the model becomes exactly $O(4)\times U(1)_X$ symmetric; adjusting the value of $M$ allows us to interpolate between these extremes and to investigate the limits to which
the custodial symmetry of the top-quark mass generating sector can be enhanced.  
When we calculate the dominant one-loop corrections to $g_{Lb}$ in our model we find, consistent with previous results \cite{Agashe:2006at}, that because $b_L$ is a $P_{LR}$ eigenstate,  $g_{Lb}$ is protected from radiative corrections in the $M\to 0$ limit and these corrections return as $M$ is switched on.   However, when we study the behavior of oblique radiative corrections as $M$ is varied, we find that in the small-$M$ limit where $g_{Lb}$ is closer to the experimental value\cite{Amsler:2008zzb}, the oblique corrections become unacceptably large. In particular, in the
$M \to 0$ limit the enhanced custodial symmetry produces a potentially sizable negative contribution to $\alpha T$.  

\section{Doublet-Extended Standard Model}
\label{sec:DESM}
\subsection{Custodial Symmetry and Z coupling}\label{cus}

The tree-level coupling of a SM fermion $\psi$ to the $Z$ boson is, 
\begin{equation}
\frac{e}{c_w s_w}\ (T^3_L-Q \sin^2\theta_W) \ Z^\mu \bar{\psi}\gamma_\mu \psi \ ,\ \ 
\label{eq:coupling}
\end{equation}
where $T^3_L$ and $Q$ are, respectively, the weak isospin and electromagnetic charges of fermion $\psi$, $e$ is the electromagnetic coupling; $c_w$ and $s_w$ are the cosine and sine of the weak mixing angle.  Because the electromagnetic charge is conserved, loop corrections to the $Z\bar{\psi}\psi$ coupling do not alter it; however, the weak symmetry $SU(2)_L$ is broken at low energies, and radiative corrections to the $T^3_L$ coupling are present in the SM.

Following Agashe \cite{Agashe:2006at} {\it et. al.}, we wish to construct a scenario in which the $T^3_L$ coupling is not subject to flavor-dependent radiative corrections.  To start, we note that the accidental custodial symmetry of the SM implies that the vectorial charge $T^3_V\equiv T^3_L+T^3_R$ is conserved 
\begin{equation}
\delta T^3_V=\delta T^3_L+\delta T^3_R=0 \ .
\label{eq:deltaT3v}
\end{equation}
This suggests a way to evade flavor-dependent corrections to $T_L^3$ itself, by adding a parity symmetry $P_{LR}$ that exchanges  $L\leftrightarrow R$.  If $\psi$ is an eigenstate of this parity symmetry and the symmetry persists at the energies of interest, then
\begin{equation}
\delta T^3_L=\delta T^3_R  \ .
\end{equation}
Now, we see that Eq. (\ref{eq:deltaT3v}) is satisfied by having the two terms on the right hand side vanish separately, rather than remaining non-zero and canceling one another.  In other words, $\delta T^3_L = 0$ and the $Z\bar{\psi}\psi$ coupling remains fixed even to higher-order in this scenario.   We will now show how to implement this idea for the $b$-quark in a toy model and examine the phenomenological consequences.

\subsection{The Model}

Let us construct a simple extension of the SM that implements this parity idea for the third-generation quarks, in order to suppress radiative corrections to the $Zb\bar{b}$ vertex.  We extend the global $SU(2)_L \times SU(2)_R$ symmetry of the Higgs sector of the SM to an $O(4)\times U(1)_X\sim SU(2)_L\times SU(2)_R\times P_{LR} \times U(1)_X$ for both the symmetry breaking and top quark mass generating sectors of the theory.  As usual, only the electroweak subgroup, $SU(2)_L\times U(1)_Y$, of this global symmetry is gauged; our model does not include additional electroweak gauge bosons.  The global $O(4)$ spontaneously breaks to $O(3) \sim SU(2)_V \times P_{LR}$ which will protect $g_{Lb}$ from radiative corrections, as above, provided that the left-handed bottom quark is a parity eigenstate:  $P_{LR} b_L = \pm b_L$.  The additional global $U(1)_X$ group is included to ensure that the light $t$ and $b$ eigenstates, the ordinary top and bottom quarks, obtain the correct hypercharges.

In light of the extended symmetry group, the relationships between electromagnetic charge $Q$, hypercharge $Y$, the left- and right-handed $T^3$ charges, and the new charge $Q_X$ associated with $U(1)_X$ are as follows:
\begin{eqnarray}
Y & = & T^3_R+Q_X \ , \label{eq:hyper} \\
Q & = & T^3_L+Y = T^3_L+T^3_R+Q_X \ . \label{eq:EM}
\end{eqnarray}
Since the $b_L$ state is supposed to correspond to the familiar bottom-quark, it has the familiar SM charges $T^3_L(b_L) = - 1/2$, and $Q(b_L) = -1/3$, and $Y(b_L) = 1/6$. Because $b_L$ must be an eigenstate under $P_{LR}$, we deduce that $T^3_R(b_L) = T^3_L(b_L) = -1/2$.  Then to be consistent with Eqs. (\ref{eq:hyper}) and (\ref{eq:EM}), its charge under the new global $U(1)_X$ must be $Q_X(b_L) = 2/3$.   Moreover, since the left-handed $b$ quark is an $SU(2)_L$ partner of the left-handed $t$ quark,  the full left-handed top-bottom doublet must have the charges $T^3_R = -1/2$ and $Q_X = 2/3$, just as the full doublet has hypercharge $Y = 1/6$.  Finally, the non-zero $T_3^R$ charge of the top-bottom doublet tells us that this doublet forms part of a larger multiplet under the $SU(2)_L \times SU(2)_R$ symmetry and it will be necessary to introduce some new fermions with $T^3_R = 1/2$ to complete the multiplet.

We therefore introduce a new doublet of fermions $\Psi  \equiv (\Omega, T^\prime)$.  The left-handed component, $\Psi_L$ joins with the top-bottom doublet $q_L \equiv (t_L^\prime, b_L)$ to form an $O(4)\times U(1)_X$ multiplet
\begin{equation}
  {\cal Q}_L = \left( {\begin{array}{*{20}c}
   {t^\prime_L } & {\Omega_L }  \\
   {b_L } & {T^\prime_L }  \\
 \end{array} } \right)
\equiv \left(\begin{array}{cc} q_L & \Psi_L \end{array}\right) \ ,
\end{equation}
which transforms as a $(2,2^*)_{2/3}$ under $SU(2)_L\times SU(2)_R\times U(1)_X$.  The parity operation $P_{LR}$, which exchanges the $SU(2)_L$ and $SU(2)_R$
transformation properties of the fields, acts on ${\cal Q}_L$ as:
\begin{equation}
P_{LR} {\cal Q}_L = - \left[ ( i \sigma_2)\, {\cal Q}_L\, (i \sigma_2) \right]^T = \left( {\begin{array}{*{20}c}
   {T^\prime_L } & {-\Omega_L }  \\
   {-b_L } & {t^\prime_L }  \\
 \end{array} } \right)
\end{equation}
exchanging the diagonal components, while reversing the signs of the off-diagonal components. Thus $t^\prime_L$ and $T^\prime_L$ are constrained to share the same electromagnetic charge, in order to satisfy Eq.~(\ref{eq:EM}).  In fact, we will later see that the $t^\prime$ and $T^\prime$ states mix to form mass eigenstates corresponding to the top quark ($t$) and a heavy partner ($T$).  The charges of the components of ${\cal Q}_L$ are listed in Table \ref{tab:charges}.

\begin{table}
{\begin{tabular}{|c||c|c|c|c||c|c|c|c|}
\hline
&$t^\prime_L$&$b_L$&$\Omega_L$ & $T^\prime_L$&$t^\prime_R$&$b_R$&$\Omega_R$ & $T^\prime_R$\\
\hline
$T^3_L$& $\frac12$ & $-\frac12$ & $\frac12$ & $-\frac12$ & $0$ & $0$ & $\frac12$ & $-\frac12$  \\
\hline
$T^3_R$& $-\frac12$ & $-\frac12$ & $\frac12$ & $\frac12$ & $0$ & $-1$ & $0 $ & $0$\\
\hline
$Q$& $\frac23$ & $-\frac13$ & $\frac53$ &$\frac23$ &$\frac23$ & $-\frac13$ & $\frac53$ &$\frac23$ \\
\hline
$Y$& $\frac16$ & $\frac16$ & $\frac76$ & $\frac76$ & $\frac23$ & $-\frac13$ & $\frac76$ &$\frac76$\\
\hline
$Q_X$& $\frac23$ &$\frac23$ &$\frac23$ &$\frac23$ &$\frac23$ &$\frac23$ &$\frac76$ &$\frac76$ \\
\hline
\end{tabular}}
\caption{\label{tab:charges} Charges of the fermions under the various symmetry groups in the model. Note that, as discussed in the text, other $T^3_R$ and $Q_X$ assignments for the $\Omega_R$ and $T^\prime_R$ states are possible.}
\end{table}

We assign the minimal right-handed fermions charges that accord with the symmetry-breaking pattern we envision:  the top and bottom quarks will receive mass via Yukawa terms that respect the full $O(4) \times U(1)_X$ symmetry, while the exotic states will have a dimension-three mass term that explicitly breaks the large symmetry to $SU(2)_L \times U(1)$.   Moreover, to accord with experiment, the  $t'_R$ and $b_R$ must have $T^3_L = 0$ and share the electric charges of their left-handed counterparts.  The top and bottom quarks will receive mass through a Yukawa interaction with a SM-like Higgs multiplet that breaks the electroweak symmetry. The simplest choice is to assign the Higgs multiplet to be neutral under $U(1)_X$; in this case, both $t'_R$ and $b_R$ share the $Q_X = 2/3$ charge of $t'_L$ and $b_L$.  Therefore, from Eqs. (\ref{eq:hyper}) and (\ref{eq:EM}), we find $T^3_R(t'_R) = 0$ (meaning that $t'_R$ can be
chosen to be an $SU(2)_R$ singlet) and $T^3_R(b_R)$ = -1 (so that $b_R$ is, minimally, part of an $SU(2)_R$ triplet if we extend the symmetry to the bottom quark mass generating sector).  Turning now to the $T'_R$ and $\Omega_R$ states, we see that they must form an $SU(2)_L$ doublet with hypercharge $7/6$ so that the Dirac mass term for $\Psi$ preserves the electroweak symmetry as desired.\footnote{This means that the $\Omega_R$ and $T^\prime_R$ states do not fill out the $SU(2)_R$ triplet to which $b_R$ belongs -- which is uncharged under $SU(2)_L$ and carries hypercharge $2/3$; other exotic fermions must play that role if we wish to extend the symmetry to the bottom quark mass generating sector.}  Finally, we choose $T^3_R(\Omega_R) = T^3_R(T^\prime_R) = 0$, which implies $Q_X = 7/6$ for both states, as the minimal choice satisfying the constraint imposed by Eq. (\ref{eq:hyper}); other choices of $T^3_R$ charge would involve adding additional fermions to form complete $SU(2)_R$ multiplets.  The charges of the fermions are listed in Table \ref{tab:charges}.

Now, let us describe the symmetry-breaking pattern and fermion mass terms explicitly.  
Spontaneous electroweak symmetry breaking proceeds through a Higgs multiplet that transforms as a $(2,2^*)_0$ under $SU(2)_L \times SU(2)_R \times U(1)_X$:
\begin{equation}
 \Phi  = \frac{1}{\sqrt{2}}\left( {\begin{array}{*{20}c}
   {v+h + i \phi^0 } & {i\sqrt{2}\ \phi^+  }  \\
   {i\sqrt{2}\ \phi^-  } & {v+h-i\phi^{0} }  \\
 \end{array} } \right)\ .
 \label{eq:higgsdef}
\end{equation}
Again, the parity operator $P_{LR}$ exchanges the diagonal fields and reverses the signs of the off-diagonal elements.  When the Higgs acquires a vacuum expectation value, the longitudinal  $W$ and $Z$ bosons acquire mass and a single Higgs boson remains in the low-energy spectrum.
The Higgs multiplet has an $O(4)\times U(1)_X$ symmetric Yukawa interaction with the top quark:
\begin{equation}
{\cal{L}}_{\rm Yukawa}= - \lambda_t \text{Tr} \left( \overline{\cal Q}_L\cdot \Phi\right) t^\prime_R \ + \rm{h.c.}\ .
\label{eq:Yuk}
\end{equation}
that contributes to generating a top quark mass. In principle, the same Higgs multiplet can also contribute to the bottom quark mass through a separate, and similarly $O(4)\times U(1)_X$ symmetric, Yukawa interaction involving the $SU(2)_R$ triplet to which $b_R$ belongs.  Since the phenomenological issues that concern us in this paper are affected far more strongly by $m_t$ than by the far-smaller $m_b$, we will neglect this and any other Yukawa interaction.

Next we break the  full $O(4)\times U(1)_X$ symmetry to its electroweak subgroup. We do so
first by gauging $SU(2)_L\times U(1)_Y$.  In addition, we wish to preserve the $O(4)$ symmetry
of the top quark mass generating sector in all dimension-4 terms, but break it 
softly by introducing a dimension-3 Dirac mass term for  $\Psi$,
\begin{equation}
{\cal{L}}_{\rm mass}= - M\ \bar{\Psi}_L\cdot\Psi_R + h.c. 
\label{eq:M}
\end{equation}
that explicitly breaks the global symmetry to $SU(2)_L\times U(1)_Y$.  Note that we therefore expect that any flavor-dependent radiative corrections to the $Zb_L\bar{b}_L$ coupling will vanish in the limit $M \to 0$, as the protective parity symmetry is restored; alternatively, as $M \to \infty$, the larger symmetry is pushed off to such high energies that the resulting theory looks more and more like the SM.

In addition to the fermions explicitly described above, a more complete version of this toy model must contain several other fermions to fill out the $SU(2)_R$ multiplet to which the $b_R$ belongs and also some spectator fermions that cancel $U(1)_Y$ anomalies. However, the  toy model suffices for exploration of the issues related to the $Zb_L\bar{b}_L$ coupling that is the focus of this paper.

\subsection{Mass Matrices and Eigenstates}

When the Higgs multiplet acquires a vacuum expectation value and breaks the electroweak symmetry, masses are generated for the top quark, its heavy partner $T$ and the exotic fermion $\Omega$ through the mass matrix:
\begin{equation}
{\cal L}_{\rm mass} = -\left(\begin{array}{*{20}c} t^\prime_L & T^\prime_L \end{array}\right)\ 
\left({\begin{array}{*{20}c} m & 0  \\ m & M \\ \end{array} } \right)
\left(\begin{array}{c} t^\prime_R \\ T^\prime_R \end{array}\right) -M \bar{\Omega}_L \Omega_R + {\rm h.c} \ ,
\end{equation} 
where
\begin{equation}
m = \frac{{\lambda _t v}}{{\sqrt 2 }} \ .
\label{eq:mtSM}
\end{equation} 
Note that the  $\Omega$ field is decoupled from the SM sector, and its mass is simply $m_\Omega = M$.
The bottom quark remains massless because we have ignored its Yukawa coupling.

Diagonalizing the top quark mass matrix yields mass eigenstates $t$ (corresponding to the SM top quark) and $T$ (a heavy partner quark), with corresponding eigenvalues
\begin{equation}
m_t^2 = \frac{1}{2}\left[1-\sqrt{1+\frac{4m^4}{M^4}}\right]M^2+m^2 \ , \ \ \ \ \ \ \ \ 
m_T^2 = \frac{1}{2}\left[1+\sqrt{1+\frac{4m^4}{M^4}}\right]M^2+m^2 \ .
\label{eq:massEv}
\end{equation}
The mass eigenstates are related to the original gauge eigenstates through the rotations:
\begin{equation}
\left(\begin{array}{c} t^\prime_R \\ T^\prime_R \end{array}\right) =
\left(\begin{array}{cc} \cos\theta_R & \sin\theta_R \\ -\sin\theta_R & \cos\theta_R \end{array}\right)
\left(\begin{array}{c} t_R \\ T_R \end{array}\right) \ ,\ \ \ \ 
\left(\begin{array}{c} t^\prime_L \\ T^\prime_L \end{array}\right) =
\left(\begin{array}{cc} \cos\theta_L & \sin\theta_L \\ -\sin\theta_L & \cos\theta_L \end{array}\right)
\left(\begin{array}{c} t_L \\ T_L \end{array}\right) \ ,
\label{eq:massEs}
\end{equation}
whose mixing angles are given by 
\begin{equation}
\sin\theta_R=\frac{1}{\sqrt{2}}\sqrt{1-\frac{1-2m^2/M^2}{\sqrt{1+4m^4/M^4}}} \ ,\ \
\sin\theta_L=\frac{1}{\sqrt{2}}\sqrt{1-\frac{1}{\sqrt{1+4m^4/M^4}}} \ .
\end{equation}
From these equations the decoupling limit $M\to\infty$ is evident: $m_t$ approaches its SM value as in Eq.~(\ref{eq:mtSM}), the $t-T$ mixing goes to zero, and $T$ becomes degenerate with $\Omega$. Conversely, in the limit $M\to 0$, the full $O(4)\times U(1)_X$ symmetry is restored and only the combination $T'_L + t'_L$ couples to $t_R$ with mass $m$.

\begin{figure}
\begin{center}
\includegraphics[width=3in]{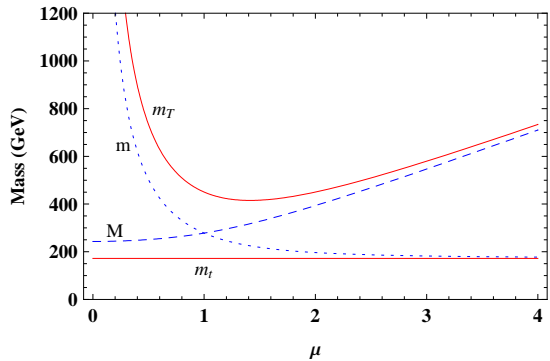}
\end{center}
\caption{The curves show the behaviors of $m$ (dotted), $M$ (dashed), and $m_T$ (upper solid) as functions of $\mu\equiv M/m$ when $m_t$ is held fixed.  The solid horizontal line corresponds to $m_t\simeq$ 172 GeV.}
\label{fig:masses}
\end{figure}

For phenomenological discussion, it will be convenient to fix $m_t$ at its experimental value and express the other masses in terms of $m_t$ and the ratio $\mu\equiv M/m$. Fig.~\ref{fig:masses} shows how $m$, $M$, and $m_T$, vary with $\mu$; the horizontal line represents $m_t$ which is being held fixed at 172 GeV.  In the limit as $\mu$ becomes large, $m\to m_t$, $m_T \sim M$ grows steadily, and the mixing angles decline toward zero; this is a physically-sensible limit that ultimately leads back to the SM.  However we see that the opposite limit, where $\mu\to 0$ can only be achieved for $m\to\infty$, which is not physically reasonable since it corresponds to taking $\lambda_t \to \infty$.  Hence, we will need to take care in talking about the case of small $\mu$.

\section{Phenomenology}
\label{sec:pheno}
\subsection{$Z$ coupling to $b_L \bar{b}_L$}
\label{sec:gLb}

We are now ready to study how the flavor-dependent corrections to the $Zb_L \bar{b}_L$ coupling behave in our toy model.  Specifically, if we write the  $Zb_L\bar{b}_L$ coupling as
\begin{equation}
\frac{e}{c_w s_w} \ \left(- \frac{1}{2} + \delta g_{Lb} + \frac{1}{3}\sin ^2 \theta _L \right)\ Z_\mu\  \bar{b}_L\gamma_\mu b_L \ ,
\label{eq:totalZbb}
\end{equation}
then all the flavor-dependence is captured by $\delta g_{Lb}$.  At tree-level, the $Z b_L \bar{b}_L$ coupling in our model has its SM value, with $\delta g_{Lb} = 0$, because the $b_L$ has the same quantum numbers as in the SM.  However, at one-loop, flavor-dependent vertex corrections arise and these give non-zero corrections to $\delta g_{Lb}$; these corrections differ from those in the SM due to the presence of vertex corrections involving exchange of $T$, the heavy partner of the top quark.

\begin{figure}
\begin{center}
\includegraphics[width=1.5 in]{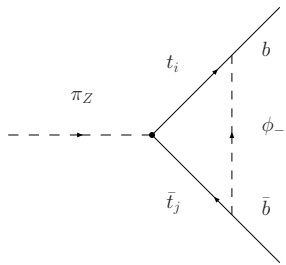}
\end{center}
\caption{One-loop vertex correction diagram for $\pi_Z \rightarrow b \bar{b}$ in our model.  The $t_{i,j}$ may be either the top quark ($t$) or its heavy partner ($T$).}
\label{fig:chibb}
\end{figure}

The calculation may be done conveniently in the ``gaugeless'' limit \cite{Barbieri:1992nz,Barbieri:1992dq,Oliver:2002up,Abe:2009ni}, in which the $Z$ boson is treated as a non-propagating external field coupled to the current $j^\mu_{3L} - j^\mu_Q \sin^2 \theta_L$.  Operationally, this involves replacing $Z_\mu$ with $\partial_\mu\phi^0/m_Z$ in the gauge current interaction, where $\phi^0$ is the Goldstone boson eaten by the $Z$:
\begin{eqnarray}
\frac{e}{c_w s_w}\ Z_\mu (j^\mu_{3L} - j^\mu_Q \sin^2 \theta_L)\ \ &\to&\ \  \frac{e}{c_w s_w m_Z}\ \partial_\mu\phi^0 (j^\mu_{3L} - j^\mu_Q \sin^2 \theta_L)\nonumber \\
& = & \frac{2}{v}\ \partial_\mu\phi^0 (j^\mu_{3L} - j^\mu_Q \sin^2 \theta_L)
\label{eq:gaugeless}
\end{eqnarray} 
The general vertex diagram shown in Fig.~\ref{fig:chibb}, will yield radiative corrections to the effective operator  $\partial_\mu\phi^0\ \bar{b}_L\gamma^\mu b_L$; that is, the expression for this diagram will include a term of the form
\beq
A\ \partial_\mu\phi^0\ \bar{b}_L\gamma^\mu b_L \ .
\label{api}
\eeq
Comparing the last three equations shows that the coefficient $A$ is proportional to the quantity we are interested in: 
\beq
\delta g_{Lb}=\frac{v}{2}\ A \ .
\eeq

\begin{figure}
\begin{center}
\includegraphics[width=3 in]{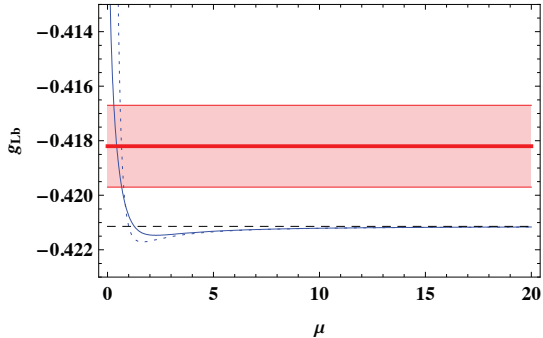}
\end{center}
\caption{The solid (blue) curve shows the DESM model's prediction for $g_{Lb}$, Eq.~(\ref{eq:dglb}). The thick horizontal line corresponds to $g_{Lb}^{ex}=-0.4182$, while the two horizontal upper and lower solid lines bordering the shaded band correspond to the $\pm 1\sigma$ deviations \cite{:2005ema}. The SM prediction is given by the dashed horizontal line. The leading-log contribution is shown by the dotted curve.}
\label{fig:gZbb}
\end{figure}

We have calculated the several loop diagrams represented by Fig.~\ref{fig:chibb} and obtain the following expression for $\delta g_{Lb}$:
\begin{eqnarray}
 \delta g_{Lb} & = & \frac{m_t^2}{16\pi^2 v^2}\left[\cos^2\theta_L\left(\cos 2\theta_L+\sin^2\theta_R\right)
+\frac{m_T^2}{m_t^2}\sin^2\theta_L \left(\cos^2\theta_R-\cos 2\theta_L\right) 
 \right. \nonumber \\
& &
\left. -\frac{m_T/m_t}{4}\sin 2\theta_L \sin 2\theta_R \right. \\
& & -\left. \frac{m_T/m_t}{2}\sin 2\theta_L\left(\frac{m_T^2/m_t^2+1}{4}\sin 2\theta_R-2\frac{m_T}{m_t}\sin 2\theta_L\right)
\frac{\log(m_T^2/m_t^2)}{m_T^2/m_t^2-1}\right] \ ,
 \nonumber \label{eq:dgZbb}
\end{eqnarray}
where the prefactor proportional to $m_t^2$ is the SM result for this class of diagram.  We expect to see  $\delta g_{Lb}$ vanish in the limit $M \to 0$ as the parity symmetry is restored; this expectation is fulfilled, since $m_t \to 0$ in this limit.  At the other extreme, for large $M$, we expect to find $\delta g_{Lb}$ take on its SM value by having the factor within square brackets approach one.  This may be readily verified if we take the equivalent limit as $\mu \to \infty$ for fixed $m_t$:
\begin{equation}
\delta g_{Lb} (\mu \to \infty)  \to \frac{m_t^2}{16\pi^2 v^2}
\left[1+\frac{\log(1/\mu^2)}{2\mu^2}+{\cal O}(1/\mu^4)\right] \ ,
\label{eq:dglb}
\end{equation}
since in this limit $\sin\theta_L \to 1/\mu^2$, $\sin\theta_R \to 1/\mu$ and $m_T^2 / m_t^2 \to \mu^2$.  In other words, we find that adjusting the value of $M$ allows us to interpolate between the SM value for $\delta g_{Lb}$ at large $M$ and the absence of a radiative correction at small $M$.  While the limit of small $\mu$ is less useful, as we mentioned earlier, for completeness we note that 
\begin{equation}
\delta g_{Lb} (\mu \to 0)  \to \frac{m_t^2}{16\pi^2 v^2}
\left[\frac{3\log(2/\mu)-1}{2} + \mu^2 \frac{6+\log(2/\mu)}{8}+{\cal O}(\mu^4)\right] .
\end{equation}
since in this limit $\sin\theta_L \to (1/\sqrt{2})(1 - \mu^2/4)$, $\sin\theta_R \to (1 - \mu^2/8)$, and $m_T^2 / m_t^2 \to 4/\mu^2$.  This growth at small $\mu$ is visible in Fig.~(\ref{fig:gZbb}).  

We now use our results to compare the value of $g_{Lb}$ in our model (as a function of $\mu$ for fixed $m_t$) with the values given by experiment and the SM, as illustrated in Fig.~(\ref{fig:gZbb}).  The experimental \cite{:2005ema}  value $g_{Lb}^{ex}=-0.4182\pm0.0015$ corresponds to the thick horizontal line; the thin (red) horizontal lines bordering the shaded band show the $\pm 1 \sigma$ deviations from the experimental value.  We calculated the SM value using ZFITTER \cite{Bardin:1999yd,Arbuzov:2005ma} with a reference Higgs mass $m_h=115\text{ GeV}$, and obtain $g_{Lb}^{SM}=-0.42114$.  This is indicated by the dashed horizontal line, and may be seen to deviate from $g_{Lb}^{ex}$ by 1.96$\sigma$.   The (solid blue) curve shows how $g_{Lb}$ varies with $\mu$ in our model; we required $g_{Lb}$ to match the SM value with $m_t = 172$ GeV and $v = 246$ GeV as $\mu \to \infty$ and the shape of the curve reflects our results for $\delta g_{Lb}$ in Eq.~(\ref{eq:dglb}).   We see that $g_{Lb}$ in our model is slightly more negative than (i.e. slightly farther from the experimental value than) the SM value for $\mu > 1$, agrees with the SM value for $\mu = 1$, and comes within $\pm 1 \sigma$ of the experimental value only for $\mu < 1$.  Given the shortcomings of the small-$\mu$ limit, this is disappointing.

\subsection{Oblique Electroweak Parameters}

The flavor-universal corrections from new physics beyond the SM can be parametrized in a model independent way using the four oblique EW parameters $\alpha S,\ \alpha T,\ \alpha \delta,\ \Delta \rho$; the first two are the oblique
paramters  \cite{Peskin:1991sw,Altarelli:1990zd,Altarelli:1991fk} for models without additional electroweak gauge bosons, while the other two incorporate the effects of an extended electroweak sector. In general, the oblique parameters are related as follows \cite{Chivukula:2004af,Barbieri:2004qk} to the neutral-current 
\begin{align}
\label{NCamplitude}
-{\cal M}_{NC} = 4\pi \alpha \frac{Q Q'}{P^2} 
& +  \frac{(T^3-s_w^2 Q) (T'^3 - s^2 Q')}
	{\left(\frac{s_w^2c_w^2}{4\pi \alpha}-\frac{S}{16\pi}\right)P^2 +	\frac{1}{4 \sqrt{2} G_F}\left(1-\alpha T + 
	\frac{\alpha \delta}{4 s_w^2 c_w^2}\right)}  \\ 
& + \sqrt{2} G_F \, \frac{\alpha \delta}{s_w^2 c_w^2}\, T^3 T'^3 
+ 4 \sqrt{2} G_F  \left( \Delta \rho - \alpha T\right)(Q-T^3)(Q'-T'^3),\nonumber
\end{align} 
and charged-current electroweak scattering amplitudes 
\beq  
- {\cal M}_{\rm CC}
  =  \frac{(T^{+} T'^{-} + T^{-} T'^{+})/2}
             {\left(\frac{s_w^2}{4\pi \alpha}-\frac{S}{16\pi}\right)P^2
             +\frac{1}{4 \sqrt{2} G_F}\left(1+\frac{\alpha \delta}{4 s_w^2 c_w^2}\right)
            }
        + \sqrt{2} G_F\, \frac{\alpha  \delta}{s_w^2 c_w^2} \, \frac{(T^{+} T'^{-} + T^{-} T'^{+})}{2},
\label{CCamplitude}
\eeq
with $P^2$ a Euclidean momentum-squared.   In the DESM we may set $\Delta \rho = \alpha T$, because the model contains no extra $U(1)$ gauge group, and $\delta=0$, because there is no extra $SU(2)$ gauge group.  We therefore work purely in terms of $\alpha S$ and $\alpha T$ from here on.  We take the origin of the $\alpha S, \alpha T$ parameter space to correspond to the SM with $m_{H}=115\text{ GeV}$; this ensures that any non-zero prediction for the oblique parameters for a Higgs of this mass arises from physics beyond the SM.  At the one-loop level, the only new contributions to $\alpha S$ and $\alpha T$ in the DESM come from heavy fermion loops in the vacuum polarization diagrams indicated in Figure \ref{fig:Pi_ij}.   We therefore expect $\alpha S$ and $\alpha T$ to be of order a few percent.\footnote{There are, in principle, additional oblique parameters such as $\alpha U$ that arise at higher order.  These will be suppressed relative to $\alpha S$ or $\alpha T$ by a factor of order $m_Z^2/m_{T}^2$; since we can see from Figure \ref{fig:masses} that $m_T > 2 m_t$, the suppression is by at least an order of magnitude and we shall neglect $\alpha U$ and its ilk from here on.} 

In this section, we will first separately derive expressions for $\alpha T$ and $\alpha S$ in DESM and see how each compares to current constraints from  \cite{Amsler:2008zzb}.  We then compare the DESM's joint prediction for $\alpha S$ and $\alpha T$ as a function of $\mu$ to the region of the $\alpha S - \alpha T$ plane that gives the best fit to existing data \cite{Amsler:2008zzb} and thereby derive a 95\% confidence level lower bound on $\mu$.  

\subsubsection{Parameter $\alpha T$}

\begin{figure}
\begin{center}
\includegraphics[width=1.5 in]{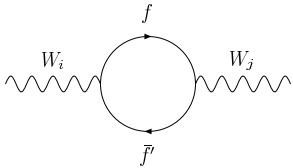}
\end{center}
\caption{Vacuum polarization diagram contributing to the oblique electroweak parameters. The indices $i,\ j=1,2,3,Q$, refer to weak (i = 1,2,3) or electromagnetic (Q) generators, while $f, f^\prime$ run over the appropriate combinations of $t$, $b$, $T$ and $\Omega$.}
\label{fig:Pi_ij}
\end{figure}

The custodial-symmetry-breaking parameter $\alpha T$ is defined as \cite{Peskin:1991sw}
\beq
\alpha T  = 
\left[ \frac{\Pi _{WW} \left( 0 \right)}{M^2_W} - \frac{\Pi _{ZZ} (0)}{M^2_Z}  \right],
\label{eq:Tdeff}
\eeq
where the contributions proportional to $g^{\mu\nu}$ in the vacuum polarization diagrams of Fig.~(\ref{fig:Pi_ij})
for the $W$ and $Z$ are labeled $\Pi_{WW}$ and $\Pi_{ZZ}$, respectively.  Each contribution sums
over various  $f \bar{f}^\prime$ pairs -- for $W$ we have $f \bar{f}^\prime =t\bar{b},\ T\bar{b},\ t \bar{\Omega},\ T \bar{\Omega}$; while for $Z$, we have $f \bar{f}^\prime=t\bar{t},\ T\bar{T},\ t\bar{T}, \Omega \bar{\Omega}, b\bar{b}, b \bar{\Omega}$.

 \begin{figure}
\begin{center}
\includegraphics[width=3 in]{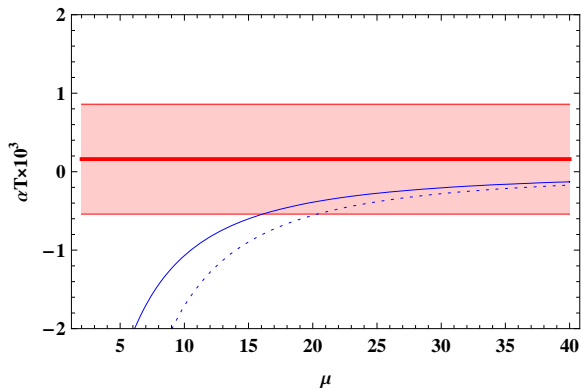} 
\end{center}
\caption{ The solid (blue) curve shows the DESM model's prediction for $\alpha T^{th}$ as a function of $\mu$. The horizontal lines show the optimal fit value of $\alpha T=0.16 \times 10^{-3}$ (thick solid line) and the relative $\pm 1\sigma$ deviations \cite{Amsler:2008zzb}  (solid lines bordering the shaded band). The line $\alpha T = 0$ corresponds to the SM value (with $m_h=115$ GeV), by definition. The leading-log contribution to $\alpha T$ is shown by the dotted curve.}
\label{fig:Drho}
\end{figure}

The analytical result for $\alpha T^{DESM}$ cannot be written in compact form;  the 
result\footnote{This is consistent with Eq. (33) in Pomarol \cite{Pomarol:2008bh} {\it et. al.}, when only the contributions from new fermions are included ($c_L = 0$).} in the limit $\mu>>1$ is:
\beq
\alpha T^{DESM}=\frac{3 m_t^2}{16\pi^2 v^2 }\left( {1 - 4\frac{{\ln \mu^2 }}{{\mu ^2 }} + \frac{{22}}{{ 3 \mu ^2 }}} \right)\,. 
\label{eq:Drho}
\eeq
One can see that, for $\mu\rightarrow \infty$, Eq.~(\ref{eq:Drho}) reproduces the leading SM result $\alpha T^{SM}(m_t)= 3 m_t^2 / (4\pi v)^2$ \cite{Peskin:1991sw}, as expected.   It interesting to note that the leading log contribution arising from the heavy states {\it reduces}\footnote{This does not violate the theorem \cite{Einhorn:1981cy, Cohen:1983fj} stating that $\Delta \rho \geq 0$ when mixing occurs only between particles of the same $T^3$ and hypercharge.  In the DESM, there is significant mixing between the $t^\prime_L$ and $T^\prime_L$ which have different $T^3$ and hypercharge values. As a result, we also expect significant GIM violation in the third generation.} the value of $\alpha T$.
This is to be expected, since the custodial symmetry is enhanced in the small-$\mu$ limit and
$\alpha T$ measures the {\it change} in the amount of isospin violation relative to the standard model. 

Subtracting the SM contribution from the top-quark, 
the numerical value of 
\begin{equation}
\alpha T^{th} = \alpha T^{DESM} - \alpha T^{SM}(m_t),
\end{equation}
as a function of $\mu$ is plotted as the solid blue curve in Fig.~(\ref{fig:Drho}); the dotted curve shows just the leading-log term (second term of Eq. (\ref{eq:Drho})).  
%Note that $\alpha T^{th} \to 0$ as $\mu \to \infty$ as we expect, since in that limit, the DESM reduces to the SM.  
The thick solid horizontal line corresponds to the best-fit value of $\alpha T = 0.16 \times 10^{-3}$ obtained by ref. \cite{Amsler:2008zzb} when setting $U = 0$; the two horizontal solid lines bordering the shaded band show the relative $\pm 1\sigma$ deviations from that central fit value.  Unlike the case of $\delta g_{Lb}$, the experimental constraints on $\alpha T$ clearly favor large values of $\mu$, closer to the SM limit.

By way of comparison, it is interesting to note that in an extra-dimensional model \cite{Carena:2006bn,Carena:2007ua} where a SM-like weak-singlet top quark was in the same $SO(5)$ multiplet as extra quarks forming a weak doublet, radiative
corrections produced an experimentally-disfavored large negative contribution to $\alpha T$ at one loop.  Given that the  $SO(5)$ multiplet in 4D includes an $SO(4)=SU(2)_L\times SU(2)_R$ bi-doublet, our results are consistent with theirs.

\subsubsection{Parameter $\alpha S$}

\begin{figure}
\begin{center}
\includegraphics[width=3 in]{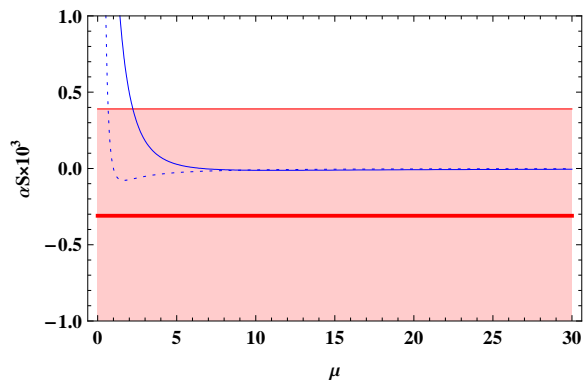}
\end{center}
\caption{The solid curve shows the DESM model's prediction for $\alpha S^{th}$ as a function of $\mu$. The horizontal lines show the optimal fit value of $\alpha S=-0.31 \times 10^{-3}$ (thick solid line) and the relative $\pm 1\sigma$ deviations (solid lines bordering the shaded band) from \cite{Amsler:2008zzb}. The line $\alpha S = 0$ corresponds to the SM value (with $m_h=115$ GeV), by definition. The leading-log contribution to $\alpha S$ is shown by the dotted curve.}
\label{fig:aS}
\end{figure}

The parameter $S$ is defined \cite{Peskin:1991sw} as
\beq
\alpha S = 16 \pi \alpha \left[ {\frac{d}{dq} \Pi _{33} \left( 0 \right) - \frac{d}{dq}\Pi _{3Q} \left( 0 \right)} \right],
\eeq
where $q$ is the gauge boson momentum. The complete expression for $\alpha S^{DESM}$ cannot be written in compact form; the limiting case where $\mu>>1$ is given by:
\beq
\alpha S^{DESM} = \frac{1}
{{6\pi }}\left( {3 + 2\ln \frac{{m_b }}
{{m_t }} + \frac{8}
{{\mu ^2 }}\left( {2 - \ln \mu } \right)} \right),
\label{eq:S}
\eeq
where we reintroduce a non-zero mass for the $b$ quark to cut off a divergence in the integral over the fermion loop momenta.  One can check that Eq.~(\ref{eq:S}) reproduces the SM result 
$\alpha S^{SM}(m_t,m_b)$ \cite{Peskin:1991sw} for $\mu \rightarrow \infty$.  Defining
\begin{equation}
\alpha S^{th} = \alpha S^{DESM}(\mu) - \alpha S^{SM}(m_t,m_b) \,,
\label{eq:delessth}
\end{equation}
we plot the result in Fig.~(\ref{fig:aS}), along with   the  value, $\alpha S=0.31\times 10^{-3}$, that provides an optimal fit to the data (for $U = 0$) and the $\pm 1\sigma$ relative deviations  \cite{Amsler:2008zzb}.  From Fig.~(\ref{fig:aS}) one can see that $\alpha S$ is within the $\pm 1\sigma$ bounds unless  $\mu < 3$; as with $\alpha T$, smaller values of $\mu$ are disfavored, though the constraint in this case is less severe.

\subsection{The $\alpha S$-- $\alpha T$ Plane}

\begin{figure}
	\begin{center}
\includegraphics[width=3 in]{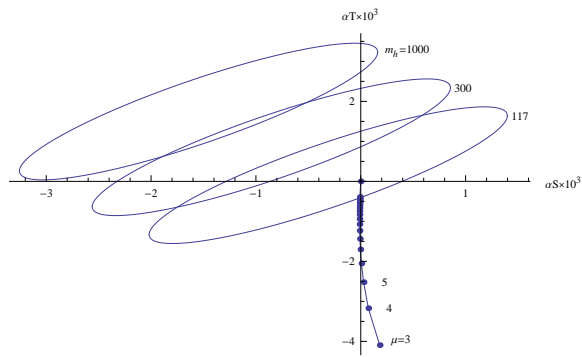}
\end{center}
\caption{The dots represent the theoretical predictions of the DESM  (with $m_h$ set to the reference value $115$ GeV), showing how the values of $\alpha S$ and $\alpha T$ change as $\mu$ successively takes on the values $3,\,4,\, 5, \,...,\,20,\,\infty$.   The three ellipses enclose the $95\%$CL regions of the $\alpha S$ - $\alpha T$ plane for the fit to the experimental data performed in \cite{Amsler:2008zzb}; they correspond to Higgs boson mass values of $m_h = 115\, {\rm GeV},\, 300\, {\rm GeV,\, and}\ 1\, {\rm TeV}$.   Comparing the theoretical curve with the ellipses shows that the minimum allowed value of $\mu$ is 20, for $m_h=115$ GeV. }
\label{fig:Ellipses}
\end{figure}

In Figure \ref{fig:Ellipses} we show the DESM predictions for  $[\alpha S^{th}(\mu), \alpha T^{th}(\mu)]$ from Eqs. (\ref{eq:delessth}, \ref{eq:Drho}) using $m_h = 115$ GeV,  and illustrating the successive mass-ratio values $\mu = 3,\, 4,\,...,20,\,\infty$; the point $\mu = \infty$ corresponds to the SM limit of the DESM and therefore lies at the origin of the $\alpha S$ - $\alpha T$ plane. 
On the same plane we also plot the elliptical curves that define the 95\% confidence level (CL) bounds on the $\alpha S$ - $\alpha T$ plane, relative to the optimal values of $\alpha S$ and $\alpha T$ found in \cite{Amsler:2008zzb}.
Using  the best-fit values \cite{Amsler:2008zzb} and corresponding $\pm 1\sigma$ deviations for $m_h = 115$ GeV, $300$ GeV, along with the correlation matrix, we obtained the approximate values appropriate to $m_h = 1$ TeV by extrapolating based on the logarithmic dependence of $\alpha S$ and $\alpha T$ on $m_h$.  To calculate the 95\% CL ellipses, 
%we used the variables $a^i$, their best-fit values, their one-sigma deviations $\sigma^i$ and the correlation matrix $\rho^{ij}$ to define:
%
%\begin{equation}
%\chi^2 = \sum_{i,j} (a^i - a_{best-fit}^i) (\sigma^2)_{ij}^{-1} (a^j - a_{best-fit}^j),\ \ \ \ \ \ \ {\rm with\ \ } (\sigma^2)_{ij} \equiv \sigma_i \rho_{ij} \sigma_j\ ,
%\end{equation}
%
%and  
we solved the equation  $\Delta\chi^2 = \chi^2 - \chi^2_{min} = 5.99$, as appropriate to the $\chi^2$ probability distribution for two degrees of freedom. 

From this figure, we observe directly that the $95\%$CL lower limit on $\mu$ for $m_h=115\text{ GeV}$ is about 20, while for any larger value of $m_h$ the DESM with $\mu\leq 20$ is excluded at 95\%CL. In other words, the fact that a heavier $m_h$ tends to worsen the fit of even the SM ($\mu \to \infty$) to the electroweak data is exacerbated by the new physics contributions within the DESM.
The bound $\mu\geq 20$ corresponding to a DESM with a 115 GeV Higgs boson also implies, at 95\%CL, that $m_{T}\geq \mu \ m_t \cong 3.4\text{ TeV}$, so that the heavy partners of the top quark would likely be too heavy for detection at LHC.  

\section{Conclusions}
\label{sec:Conclusion5}

We have introduced the doublet-extended standard model (DESM) as a simple realization of the idea \cite{Agashe:2006at} of using custodial symmetry to protect the $Zb_L\bar{b}_L$ coupling ($g_{Lb}$) from receiving large radiative corrections.  In this toy model, all terms of dimension-4 in the top-quark mass generating sector
obey a global $O(4)\times U(1)_X$ symmetry, which includes a parity symmetry protecting $g_{Lb}$ from radiative corrections.  That global symmetry is softly broken to its $SU(2)_L \times U(1)_Y$ subgroup by a Dirac mass term for the extra fermion doublet that incorporates the heavy partner of the top quark.  Varying the size of this Dirac mass $M$ allows the model to interpolate between the $O(4)\times U(1)_X$-symmetric case ($M=0$) in which $\delta g_{Lb} = 0$ and the SM-like case ($M \to \infty$) in which the one-loop corrections to $g_{Lb}$ are as in the SM, and
enabled us to investigate the degree to which
the custodial symmetry of the top-quark mass generating sector can be enhanced.  
By comparing the predictions of the DESM with experimental constraints on the oblique parameters $\alpha S$ and $\alpha T$ from \cite{Amsler:2008zzb}, we found the DESM to be consistent with experiment only for $\mu>20$ at 95\%CL, with a Higgs mass $m_h=115$ GeV. The bound on $\mu$ translates into the 95\% CL lower bound of $3.4\text{ TeV}$ on the masses of the extra quarks -- placing them out of reach of the LHC. This result demonstrates 
that electroweak data strongly limits the
amount by which the custodial symmetry of the top-quark mass generating sector can be enhanced relative
to the standard model. 

While we cannot discuss this here, the toy model reproduces the behavior
of  extra-dimensional models in which
the extended custodial symmetry is invoked to control the size of additional contributions to $\alpha T$ and
the $Zb_L\bar{b}_L$ coupling \cite{Carena:2006bn,Carena:2007ua},  while leaving the standard model contributions essentially unchanged. Finally, we also note that our results are also consistent with a general effective field-theory analysis,\cite{Pomarol:2008bh}
which confirms that the toy DEWSB model illustrates the electroweak physics operative in a broad class of models.

{\bf Acknowledgments:}
This work was supported in part by the US National Science Foundation under grant PHY-0354226.  RSC and EHS acknowledge the support of the Aspen Center for Physics and the Institute for Advanced Study
 where part of this work was completed.

%\bibliographystyle{ws-procs975x65}
%\bibliography{ws-pro-sample}

\end{document}